\documentclass[12pt]{article}
\usepackage{cite,graphicx,amsmath,amssymb}

\newcommand{\be}{\begin{equation}}  
\newcommand{\ee}{\end{equation}}  
\newcommand{\bea}{\begin{eqnarray}}  
\newcommand{\eea}{\end{eqnarray}}  
\newcommand{\ol}[1]{\overline{#1}}
\newcommand{\hc}{+\,\mathrm{h.c.}}

\begin{document}

\thispagestyle{empty}
\vspace*{.2cm}
\noindent
HD-THEP-06-08 \hfill 23 May 2006

\vspace*{1.5cm}

\begin{center}
{\Large\bf SUSY breaking mediation by throat fields}
\\[2.5cm]
{\large F.~Br\"ummer, A.~Hebecker, and M.~Trapletti}\\[.5cm]
{\it Institut f\"ur Theoretische Physik, Universit\"at Heidelberg,
Philosophenweg 16 und 19, D-69120 Heidelberg, Germany}
\\[.5cm]
{\small\tt (\,f.bruemmer, a.hebecker,} {\small and}
{\small\tt m.trapletti@thphys.uni-heidelberg.de)}
\\[2.0cm]

{\bf Abstract}
\end{center} 

\noindent 
We investigate, in the general framework of KKLT, the mediation of 
supersymmetry breaking by fields propagating in the strongly warped region 
of the compactification manifold (`throat fields'). Such fields can couple 
both to the supersymmetry breaking sector at the IR end of the throat and
to the visible sector at the UV end. We model the supersymmetry breaking 
sector by a chiral superfield which develops an $F$ term vacuum expectation 
value (also responsible for the uplift). It turns out that the
mediation effect of vector multiplets propagating in the throat can compete 
with modulus-anomaly mediation. Moreover, such vector fields are naturally 
present as the gauge fields arising from isometries of the throat (most 
notably the SO(4) isometry of the Klebanov-Strassler solution). Their 
mediation effect is important in spite of their large 4d mass. The latter is 
due to the breaking of the throat isometry by the compact manifold at the UV 
end of the throat. The contribution from heavy chiral superfields is found 
to be subdominant.

\newpage
\section{Introduction}
The KKLT scenario~\cite{Kachru:2003aw} suggests a way in which all moduli 
of a string compactification could be stabilized in a de Sitter vacuum with 
small cosmological constant. The construction is based on the supergravity 
approximation to type IIB superstring theory. The complex structure moduli 
of the internal manifold and the dilaton are stabilized 
by fluxes (see e.g.~\cite{Grana:2005jc} and references therein). There is 
always at least one K\"ahler modulus $T$ which does not appear in the flux 
superpotential and which KKLT take to be stabilized nonperturbatively (e.g.
by gaugino condensation). This results in a supersymmetric AdS vacuum. SUSY 
may be broken and the cosmological constant uplifted to a positive value by 
introducing $\ol{\rm D3}$ branes. 

The flux-stabilized compactification geometry for type IIB models was worked 
out in~\cite{Giddings:2001yu}, where it was found that fluxes lead to a warped 
product structure between the four non-compact spacetime dimensions and the 
internal manifold. Strongly warped regions or `throats' occur naturally, 
with a large hierarchy of scales between the bottom of the throat (`IR 
end') and the weakly warped remainder of the internal space (`UV end' or 
`compact manifold'). This hierarchy is crucial for the KKLT construction 
since the $\ol{\rm D3}$ branes are dynamically confined to the IR end of the 
throat, where their contribution to the vacuum energy density is maximally
redshifted. Thus it is possible to start with a pre-uplift AdS vacuum at 
parametrically large volume, and still (nearly) cancel its parametrically
small cosmological constant.

In this paper we study the mediation of SUSY breaking between the bottom 
of the throat and the UV region. Taking the visible sector to be localized 
in the UV and the SUSY breaking sector in the IR, we focus on the mediation 
effects of fields propagating in the throat and coupling to both sectors. 
In KKLT, supersymmetry breaking is induced by the $\ol{\rm D3}$ branes, which 
represents a hard breaking from the point of view of 4d supergravity. The 
systematic analysis of soft terms in this setup was pioneered 
in~\cite{Choi:2005ge}, employing non-linearly realized SUSY for the 
description of the uplifting sector in the throat.\footnote{
See 
also \cite{Choi:2004sx}. See \cite{Endo:2005uy}
for some other work on soft 
terms in KKLT, and \cite{Camara:2003ku} for a selection of earlier papers on 
soft terms in type IIB flux compactifications.
}
Here, we will instead introduce a dynamical chiral superfield $X$, confined 
to the bottom of the throat. The superpotential and K\"ahler potential of 
$X$ are assumed to ensure a non-zero vacuum expectation value of its 
auxiliary component $F_X$ (see e.g.~\cite{Luty:2002hj,Lebedev:2006qq} for
related examples involving $F$ term uplifts). It will turn out that much of 
the SUSY breaking dynamics as seen by the visible sector is independent of the
details of the $X$ Lagrangian. Our SUSY breaking $X$ sector may be seen in two 
ways: On the one hand, it can model the $\ol{\rm D3}$ brane sector (including, 
in the limit of an extremely heavy $X$, much of what can be done using 
non-linearly realized SUSY). On the other hand, it can be taken seriously in 
its own right, expecting that, in the future, a stringy realization of such 
a sector will be found and used to replace the $\ol{\rm D3}$ brane sector of 
KKLT. For example, it could be realized by branes at singularities at the 
bottom of the throat \cite{Berenstein:2005xa}.

Our main result is that, under certain quite natural conditions, the 
sequestering assumption (see~\cite{Randall:1998uk} and the critical 
discussion of~\cite{Anisimov:2001zz}) is generically violated by effective 
5d gauge fields representing the SO(4) isometry of the Klebanov-Strassler 
throat~\cite{Klebanov:2000hb,Herzog:2001xk}. More specifically, we assume 
that our SUSY-breaking field $X$ is charged under the isometry of the throat.
This is natural given that, for example, $\ol{\rm D3}$ branes at the bottom 
of the throat break this isometry. Furthermore, we assume that the compact 
space at the UV end of the throat breaks the isometry softly. Again, this is 
natural since one can clearly imagine a compact manifold respecting (part of) 
the throat isometries and view the actual Calabi-Yau manifold of a realistic
model as a deformation thereof. Under these conditions, it is easy to 
see that the massive vector fields originating in the isometry develop 
a $D$ term and that this $D$ term induces SUSY-breaking scalar masses in 
the UV sector of the model which can compete with those of the sequestered 
case.

We do not consider this a negative result for the following reason: 
Our findings suggest that the above effect of `vector mediation' is the 
only one competing with mixed modulus-anomaly mediation. Thus, it is 
conceivable that, in specific models, the impact on soft scalar masses will 
turn out to be small or, more interestingly, will be large and calculable. 
Clearly, this will require a better understanding of both the SUSY breaking 
sector at the bottom of the throat as well as the geometry of the compact 
space.

While this work was being finalized, Ref.~\cite{Choi:2006bh} appeared, which 
uses the formalism of non-linear realizations and focuses on the effect 
of anomalous U(1) gauge symmetries. Nevertheless, in its discussion of 
sequestering it has a significant overlap with this analysis. Our
perspective differs in that we identify throat vector fields (which are 
also discussed in~\cite{Choi:2006bh}) as an intrinsic feature of the KKLT 
setup, forced upon us by the symmetry of the supergravity solution. 
Furthermore, our dynamical description of the SUSY breaking sector based on 
the chiral superfield $X$ allows us to specify its effect on isometry vector 
fields in a very direct and physical way. We will comment on some additional 
fine points in which our analyses differ as we go along. 

The present paper is organized as follows. We begin with a streamlined 
discussion of the basic KKLT setup and its SUSY breaking dynamics in 
Sect.~2. The use of the chiral compensator formalism and the realization of 
the SUSY breaking and uplifting sector as conventional $F$ term breaking make 
the discussion of energy scales and sequestering particularly transparent. 

In Sect.~3 we introduce, motivated by the isometry of the Klebanov-Strassler 
throat, bulk vector fields and calculate the $D$ terms they acquire if 
the SUSY-breaking chiral superfield $X$ is charged. It turns out that their 
effect on the scalar masses of Standard Model (or other chiral) superfields 
in the UV sector is potentially large.

For completeness, we analyse in Sect.~4 the effect of throat fields whose 
zero modes are 4d chiral superfields. Assuming that those chiral superfields
acquire large masses due to some dynamics in the UV region and that they 
do not mix with $T$ at the perturbative level, we find that their effect 
is small and sequestering is respected.

Section~5 contains our conclusions and a brief discussion of open issues.
Two technical calculations related to a concrete SUSY breaking model and to 
the mediation effect of massive chiral superfields are given in the Appendix.

\section{The minimal scenario}

Following~\cite{Kachru:2003aw}, we consider a type IIB compactification with 
all complex structure moduli and the dilaton stabilized by fluxes at a 
supersymmetric minimum, at which the superpotential is $W_0$. We assume 
that there is only a single K\"ahler modulus $T$ with no-scale kinetic 
function and K\"ahler potential
\be\label{noscale}
\Omega=-(T+\ol{T})\qquad\mbox{and}\qquad K=-3\,\log(-\Omega)=-3\,
\log(T+\ol{T})\,.
\ee
It is stabilized, e.g., by gaugino condensation, such that the total 
superpotential becomes
\be\label{superpot}
W=W_0 + e^{-T}\,.
\ee 
We suppress ${\cal O}(1)$ numerical coefficients here and below and 
work in units where $\alpha'\sim 1$. 

The scalar potential (in a frame where $\Omega$ is the coefficient of the 
Einstein-Hilbert term) is most easily obtained from the standard 
$D=4,\;{\cal N}=1$ Lagrangian
\be\label{sugralagrangian}
{\cal L}=\int d^4\theta \,\ol{\varphi}\varphi\,\Omega+\left(\int d^2\theta\, 
\varphi^3\,W\hc\right)\,,
\ee
where $\varphi=1+\theta^2F_\varphi$ is the chiral 
compensator~\cite{Cremmer:1982en}.

With the above kinetic function and superpotential (and using the same 
symbol for a chiral superfield and its lowest component), the 
scalar-potential-part of Eq.~\eqref{sugralagrangian} is
\be\label{boslagrangian}
{\cal L}\supset-(T+\ol{T})\,|F_\varphi|^2-(F_T F_{\ol{\varphi}}\hc)+
\Bigl[\bigl(3(W_0+e^{-T})F_{\varphi}-e^{-T}F_T\bigr)\hc\Bigr].
\ee
The resulting equations of motion are
\begin{align}
F_{\ol{\varphi}}:\quad& -(T+\ol{T})F_\varphi-F_T+3(W_0+e^{-\ol{T}})=0,
\label{fphieom}\\
F_{\ol{T}}:\quad&-F_\varphi-e^{-\ol{T}}=0,
\label{fteom}\\
\ol{T}:\quad&-|F_\varphi|^2-3e^{-\ol{T}}F_{\ol{\varphi}}
+e^{-\ol{T}}F_{\ol{T}}=0.\label{teom}
\end{align}
Taking $W_0$ to be parametrically small (which may be justified by the 
exponentially large number of flux choices), it is easy to see that 
the above equations are solved for
\be
F_T\sim F_\varphi \sim e^{-T} \sim W_0\,.
\ee
Note that here and below we focus on parametrically small factors 
$\sim e^{-T}$ but ignore factors $\sim 1/T$ (which are strictly speaking 
also parametrically small, but to a much lesser degree). The vacuum energy 
density is negative and $\sim W_0^2$. 

At this point, it is crucial to recall that a solution of the above 
equations does not represent a true vacuum of the model unless the 
curvature scalar (which is multiplied by $\Omega$) vanishes. This 
shortcoming will now be corrected. 

Assume that, due to the fluxes, the compactification manifold has 
developed a strongly warped region or throat. We normalize the warp factor 
such that it is ${\cal O}(1)$ at the UV end (and throughout the compact 
space) and equals $\omega$ (with $\omega\ll 1$) deep in the IR. Thus, mass 
scales at the bottom of the throat are redshifted by a factor $\omega$.

Now let us add a SUSY-breaking sector in the throat, i.e.~some physical 
degrees of freedom localized at the bottom of the throat which can, in 4d 
language, be described by a chiral superfield $X$. We assume that $X$ is 
sequestered from all the light fields on the compact space in the
UV\footnote{Alternatively, a dynamical SUSY-breaking can be introduced via
$D$ terms~\cite{Burgess:2003ic} or non--sequestered 
$F$ terms~\cite{Lebedev:2006qq}.}. 
(At the  present stage of our analysis, the only relevant light field
is the universal K\"ahler modulus $T$.)
As discussed in some detail in~\cite{Brummer:2005sh}, 
this setup can be viewed as a Goldberger-Wise stabilized Randall-Sundrum 
model with $T$ being a no-scale field localized at the UV brane 
(see~\cite{Giddings:2005ff} for general analysis of deformations of warped 
models). Thus, the sequestering assumption can indeed be made and the $X$- 
or uplifting sector leads to corrections of the type~\cite{Luty:2000ec}
\bea\label{uplift}
\Omega_{\rm up}&=&\Omega+\omega^2\,\Delta\Omega(X,\ol{X}) \nonumber\\
W_{\rm up}&=&W+\omega^3\,\Delta W(X)=W_0+e^{-T}+\omega^3\,\Delta W(X)\,.
\eea
Here the warp factor dependence (which follows on dimensional grounds) has
been given explicitly. Therefore the functions $\Delta W$ and $\Delta\Omega$ 
are naturally of the order of the string scale (i.e. they are ${\cal O}(1)$ 
in our units and contain no small or large parameters). 

Note that we could in principle have performed a K\"ahler-Weyl rescaling 
by $T^\alpha$ before imposing sequestering, which would 
correspond to using 
\be
\Omega=-(T+\ol T)(T\ol T)^\alpha,\qquad W=T^{3\alpha}(W_0+e^{-T})
\ee
in Eq.~\eqref{uplift}. The value of $\alpha$ can be fixed by requiring the uplift energy
to scale as \mbox{$1/(T+\bar{T})^2$}~\cite{Kachru:2003sx}. It is easily checked that $\alpha=0$ 
is the correct value, which shows that, in Eq.~\eqref{uplift}, we should indeed use the 
unrescaled form of $\Omega$ and $W$ as given in Eqs.~\eqref{noscale} and \eqref{superpot}. 


Neglecting for the moment the influence of $F_\varphi$ on the $X$ sector
(this will be easy to justify a posteriori), the equation of motion for 
$F_{\bar{X}}$ reads
\be
\omega^2\Delta\Omega_{X\bar{X}}F_X+\omega^3\Delta \bar{W}_{\bar{X}}
=0\,,
\ee
where the indices of $\Delta W$ and $\Delta\Omega$ denote partial 
derivatives. 

Obviously, if $\Delta W$ and $\Delta\Omega$ are such that,
in the absence of warping, $F_X$ would break SUSY at the string scale, then
\be
F_X\sim\omega
\ee
in the present context. The vacuum energy density induced by the $X$ sector 
is $\sim \omega^4$, with comparable contributions coming from $\Delta W$ and 
$\Delta\Omega$. 

It now becomes clear that to uplift the previously found negative vacuum 
energy density $\sim W_0^2$ to a realistic positive value (i.e. to zero, for 
all practical purposes), we need $W_0\sim \omega^2$. Thus, there is in fact 
only one small parameter in the model, which we choose to be $\omega$. It
is also immediately clear that, in this situation, the influence of the 
$X$ sector on the previously found solution for $F_\varphi$ (and hence on
$T$ and $F_T$) is of higher order in $\omega$. Thus, Eqs.~(\ref{fphieom}) - 
(\ref{teom}) continue to be the right equations to solve. The $X$ sector 
simply adds the necessary positive vacuum energy to promote the solutions
of these equations to a physical vacuum with 
\be 
F_T\sim F_\varphi \sim W_0 \sim \omega^2 \qquad\mbox{and}\qquad F_X \sim 
\omega\,.
\ee
This lies at the basis of `mixed modulus-anomaly
mediation'~\cite{Choi:2005ge}\footnote{In detail, the small hierarchy
between $F_T$ and $F_\varphi$ based on the $\beta$-function coefficient of
the condensing gauge group and ignored in this paper also plays an
important r\^ole.}.
Recall that, due to sequestering, $F_X$ has no direct effect on soft terms
in the visible sector.

\section{Vector superfields}

In this section we study the effect of a massive vector superfield
on sequestering. Such a field can emerge from an isometry of the throat, 
which becomes a gauge symmetry in the corresponding 4d field theory. If the 
isometry is not a symmetry of the entire internal manifold (in particular,
of the UV end), this gauge symmetry is spontaneously broken, and the
gauge field acquires a UV-scale mass. 

The prime example of a throat is the warped deformed conifold of Klebanov 
and Strassler~\cite{Klebanov:2000hb}. Its full 10d geometry is a deformation 
of AdS$_5\times T^{1,1}$, where $T^{1,1}$ is a 5d compact Einstein space 
with isometry $G=\;$SU(2)$\,\times\,$SU(2)$\,\times\,$U(1). The massless 5d 
spectrum of type IIB supergravity compactified on $T^{1,1}$ contains seven 
vector multiplets from the seven generators of $G$~\cite{Ceresole:2000jd}. 
In the warped deformed conifold the U(1) is broken, and the isometry group
is SU(2)$\,\times\,$SU(2)$\;=\;$SO(4)~\cite{Herzog:2001xk}. Hence the 5d 
solution should contain six massless vector multiplets. This solution can 
be interpreted as the bulk of a Randall-Sundrum-like model, with the 
Calabi-Yau playing the role of the UV brane. Since Calabi-Yaus admit no 
isometries, the zero modes of the above vector multiplets acquire a large 
mass which, from the point of view of the Randall-Sundrum model, can be 
ascribed to a UV-brane-localized mass operator. 

Independently of the above UV-scale breaking of the SO(4) gauge symmetry, we 
assume that the SUSY breaking sector at the bottom of the throat by itself 
also breaks this symmetry. In particular, a $\ol{\rm D3}$ brane at the bottom 
of the throat already breaks part of the isometry. Clearly, 
as far as the mass of the 4d vector states is concerned, this IR-scale 
breaking can not compete with the UV-scale breaking discussed earlier. 
We model this situation in the following by assuming that our SUSY-breaking 
sector contains fields charged under the gauge symmetry, but ignoring any 
possible symmetry breaking $A$ term vevs from this sector.\footnote{
The
model developed in the last two paragraphs differs from the 5d massive 
vector fields discussed in~\cite{Choi:2006bh}, yet the following technical 
analysis will lead to very similar results.
}

To derive the main qualitative SUSY breaking effects of the above setup,
we introduce a single 4d vector superfield $V$ (although the actual symmetry 
is non-abelian and several such fields are expected). $V$ has the following 
component expansion:
\be
V=C+\theta\sigma^\mu\ol{\theta} A_\mu+
\frac{1}{2}\left(F_V\theta\theta\hc\right)+
\frac{1}{2}\theta\theta\ol{\theta}\ol{\theta} D+\;{\rm fermions}.
\ee
Here $F_V$ is complex, while $A_\mu,C,D$ are real. The UV-brane symmetry 
breaking (or non-linear realization of the gauge symmetry) is modelled by 
simply giving this vector superfield a string-scale mass term. A massive 
vector superfield can give rise to soft terms in two ways: it may develop 
$F$ or $D$ terms in the vacuum.\footnote{
Note 
that for a massless vector superfield $V$, $F$ terms are unphysical because 
the $\theta^2$-components of $V$ can be gauged away using Wess-Zumino gauge. 
This is no longer the case when $V$ is massive.
}

The dominant effect is easy to guess: Focus on the term $CD$ (coming from the 
superfield mass term $\sim V^2$) and the term $\omega^2C|F_X|^2$ (coming 
from the gauge coupling $\omega^2\bar{X}VX$ motivated above). Varying these 
terms with respect to $C$ one immediately finds 
\be 
D\sim \omega^2|F_X|^2 \sim \omega^4\,,
\ee
which induces scalar masses $\sim\omega^4$ for standard model fields $Q$ in 
the visible sector if a gauge coupling $\bar{Q}VQ$ exists. The presence of 
such a gauge coupling represents, of course, a crucial assumption to which 
we will return at the end of this section. We note that the $D$ term 
contribution to the vacuum energy density is 
negligible compared with $|F_X|^2$, which is responsible for the uplift.

To derive the above in more detail, we start with the Lagrangian
\bea
{\cal L}&=&\int d^4\theta\,\ol{\varphi}\varphi\left[\Omega(T,\ol{T})+\omega^2
\Delta\Omega(X,e^V\ol{X})+V^2\right]\\
&&+\left(\int d^2\theta\,\left[\varphi^3\left\{W(T)+\omega^3\Delta W(X)
\right\}+\frac{1}{4}{\cal W}^\alpha{\cal W}_\alpha\right]\;\hc\right),
\eea
where ${\cal W}_\alpha$ is the field strength chiral superfield corresponding
to $V$. The most relevant terms in the Lagrangian are the mass term for $V$,
the gauge-kinetic term, as well as the terms
\be
\Omega=-(T+\ol{T})+\omega^2\,X(1+V)\ol{X}+\ldots\qquad W=W_0+e^{-T}+\ldots
\ee
(as before we suppress any coefficients that are generically of order one).
In components, the mass term contributes
\be
\varphi\ol{\varphi}V^2\rvert_{\theta^4}=CD+F_VF_{\ol{V}}+A_\mu A^\mu
+C\,(F_\varphi F_{\ol{V}}\hc)+C^2F_\varphi F_{\ol{\varphi}},
\ee
and the gauge kinetic term gives
\be
{\cal W}^\alpha{\cal W}_\alpha\rvert_{\theta^2}=-\frac{1}{2}
F_{\mu\nu}F^{\mu\nu}
+D^2\,.
\ee
{}From the coupling of the gauge field to the SUSY breaking field $X$ we get
\be\begin{split}
\ol{\varphi}\varphi XV\ol{X}\rvert_{\theta^4}=&
F_{\ol{\varphi}}F_\varphi X C \ol{X}+F_{\ol{\varphi}}F_X C \ol{X}\hc
+\frac{1}{2} F_{\ol{\varphi}}F_V X\ol{X}\hc\\&+
F_XF_{\ol{X}} C+\frac{1}{2} F_XF_{\ol V}\ol{X}\hc
+\frac{1}{2}XD\ol{X}.
\end{split}\ee
For $X=0$ in the vacuum (we can define $X$ such that any nonvanishing vev 
is already absorbed in the gauge symmetry breaking mass term for $V$), 
the equations of motion are
\begin{align}
C:\quad& D+(F_\varphi F_{\ol{V}}\hc)+2CF_\varphi F_{\ol{\varphi}}
+\omega^2 F_X F_{\ol{X}}=0,\\
F_{\ol{V}}:\quad& F_V+2CF_\varphi=0,\\
D:\quad& C+D=0,
\end{align}
giving
\be
C\sim\omega^4,\quad D\sim\omega^4,\quad F_V\sim\omega^6.
\ee

It is obvious that $F_V$ is irrelevant for SUSY breaking mediation. $D$,
however, will contribute significantly, because a possible coupling 
$\sim Qe^V\ol{Q}$ to the visible sector will clearly induce soft scalar 
masses $m^2\sim D\sim\omega^4$. This is just the same order of magnitude as
we get from mixed modulus-anomaly mediation, so `vector mediation' will 
compete with these effects. Of course, this can also be easily seen by 
focusing on the couplings $\sim \omega^2Xe^V\ol{X}$ and $\sim Qe^V\ol{Q}$ 
and integrating out the heavy vector. The induced operator 
$\omega^2 X\ol X Q\ol Q$ provides soft masses 
$m^2\sim\omega^2|F_X|^2 \sim \omega^4$. 

Note that the sign of the operator $\omega^2 X\ol X Q\ol Q$ depends on
the relative sign of the $X$ and $Q$ charges. In the case that it 
is negative, the contribution to the $Q$ mass term is tachyonic. However,
this effect need not lead to dangerous instabilities since other 
contributions, e.g.~from mixed modulus-anomaly mediation, are of the same 
order of magnitude and can thus cancel the vector-mediated contributions. 
On the other hand, it can be avoided altogether by a suitable choice of 
charges --- the usual constraints on gauge charges from anomaly cancellation 
do not apply here, because $V$ is massive and the gauge symmetry is broken.
In summary, we can always arrange for $Q$ to be non-tachyonic, so that
in particular the $D$ term in the vacuum is nonzero.

Finally, we want to argue in favour of the assumed coupling of the visible
sector fields $Q$ to $V$. Imagine, for example, that a stack of D branes 
wrapped appropriately on the Calabi-Yau space in the UV gives rise to a
higher-dimensional gauge theory. Let the scalars of this SUSY gauge theory 
be the superpartners of standard model fermions (`matter from gauge'). Since 
the throat isometry may, among other effects, be broken in the UV by the 
position of this brane stack, it is natural to expect that the matter
superfields are charged under $V$. This provides a strong motivation for 
the coupling $\bar{Q}e^{V}Q$ used above.

\section{Chiral superfields}
Consider now a chiral superfield $Y$ with a string-scale mass term, such 
as might be produced by flux stabilization. 
We will show that, even if $Y$ has direct couplings to both the visible and 
the SUSY breaking sector, sequestering is not violated to leading order, as 
the contributions of $Y$ to SUSY breaking mediation are subdominant. We will
estimate the $F$ term of $Y$ in the vacuum, since it may give SUSY breaking
soft masses to visible sector fields via terms like $\ol{Y}Y\ol{Q} Q$.

Suppressing ${\cal O}(1)$ coefficients, the dominant terms in the kinetic
function and superpotential are 
\be\begin{split}
\Omega&=-(T+\ol{T})+Y\ol{Y}+\omega^2\,(X\ol{Y}\hc)+\ldots\\
W&=W_0+Y^2+(1+Y)e^{-T}+\omega^3\,XY+\ldots\end{split}\,.
\ee
Since we imagine that $Y$ contains fields propagating in the throat, we have
allowed for the strongest possible couplings to the $X$ sector. Furthermore, 
since $Y$ does not represent a modulus of the fluxed Calabi-Yau, we have 
allowed for an unsuppressed mass term $\sim Y^2$ but excluded any 
leading-order linear term in $Y$ or a mixing of $Y$ and $T$. However, once 
non-perturbative effects (e.g. gaugino condensation) are incorporated, the 
clear separation between $Y$ and $T$ may be blurred, which motivates us to 
include the term $\sim Ye^{-T}$, as an example for such effects.\footnote{
This 
is a slight generalization of the otherwise similar analysis 
of~\cite{Choi:2006bh}
}
Note that we could have replaced $Y\ol{Y}$ by $(T+\ol{T})Y\ol{Y}$ 
without affecting the results of the following analysis. 

Since $F_\varphi$ would by itself not generate a non-zero $F_Y$, we will
neglect its influence for the moment. Afterwards we will show that the
backreaction of $Y$ on $F_\varphi$ is indeed negligible, hence this ansatz
is fully self-consistent. 

We obtain the following $Y$- and $F_Y$-dependent terms in the bosonic 
Lagrangian:
\be
\begin{split}{\cal L}\;\supset\;&F_YF_{\ol{Y}}+\omega^2\,(F_XF_{\ol{Y}}\hc)
+\left[2Y F_Y+F_Ye^{-T}-Y e^{-T}F_T\hc\right]
\\
&+\omega^3(F_X Y+X F_Y\hc)\,.\end{split}
\ee
Recall that $e^{-T}\sim\omega^2$ by assumption. This leads to the 
equation of motion for $Y$
\be\label{yeom}
2 F_Y-\omega^2 F_T+\omega^3F_X=0\,,
\ee
hence
\be\label{fy}
F_Y\sim\omega^2 F_T\sim\omega^3 F_X\sim\omega^4.
\ee
The equation of motion for $F_{\ol{Y}}$ reduces to
\be
F_Y+\omega^2 F_X+2 \ol{Y}+\omega^2+\omega^3\ol{X}=0\,,
\ee
thus
\be
Y \sim\omega^2.
\ee

To ensure that this estimate is correct, we now need to prove that there
are no contributions to $F_\varphi$ and $F_T$ of order $\omega^2$. This is 
fairly obvious, however, since what we are adding to the Lagrangian by 
including $Y$ is, in the vacuum, suppressed by sufficiently high powers of
$\omega$. For example, we can check that Eq.~\eqref{fteom}, the equation of 
motion for $F_{\ol{T}}$, now becomes 
\be
F_\varphi+(1+Y)e^{-T}=0\,,
\ee
inducing a negligible correction to $F_\varphi\sim\omega^2$. (This remains 
correct if $Y\ol{Y}$ is replaced by $(T+\ol{T})Y\ol{Y}$. Similarly, it is
easy to check that the vacuum values of $T$ are not affected at leading 
order in $\omega$. 

In summary, we have seen that throat fields which are described by heavy 
chiral superfields in the 4d effective theory cannot contribute sizeably 
to SUSY breaking mediation because their $F$ terms are always subdominant 
compared to $F_\varphi$ and $F_T$.

We will give an alternative (and somewhat more general) derivation of the 
warp factor dependence in appendix B, using standard supergravity relations.

\section{Conclusions}

In this paper we have studied SUSY breaking mediation in the KKLT setup, in
particular the contributions from vector and chiral fields propagating in
the throat. Vector fields are naturally present in warped throat
constructions. For instance, the currently best-understood example of 
a warped throat, the Klebanov-Strassler solution, has an SO(4) global 
symmetry acting on the compactification manifold and the flux background, 
hence there will be six vector fields characterizing the corresponding gauge 
symmetry in the compactified theory. These will become massive because at 
the UV end of the throat the symmetry will in general be broken. We find 
that, despite their string-scale mass term, they can make a
contribution to SUSY breaking mediation which is equally important as that
from modulus-anomaly mediation. In detail, from earlier 
analyses~\cite{Choi:2005ge} one can see that the soft masses induced by 
modulus-anomaly mediation are of the order $m^2\sim\omega^4M_P^2$, where 
$\omega\ll 1$ is the relative redshift between the UV and IR ends of the 
throat. This is the same order of magnitude as we find for additional
vector fields. Therefore in any realistic scenario it will be essential to 
take `vector mediation' into account as one of the leading effects. By 
contrast, the contribution from heavy chiral superfields is suppressed by 
another factor $\omega^4$, so these are truly irrelevant for communicating 
SUSY breaking.

It would be desirable to have a concrete model in which the fields 
propagating in the throat are identified in terms of the supergravity 
solution and SUSY breaking mediation effects can be explicitly calculated. 
A step in this direction would be to understand properly the 5d SUSY 
description of the Klebanov-Strassler throat. Such an intermediate-scale 
5d picture, applicable before going to 4d, could capture the geometric 
sequestering properties of the model, and yet be much simpler to analyze 
than the full 10d system (see e.g.~\cite{Brummer:2005sh,Gherghetta:2006yq} 
for 5d approaches to the Klebanov-Strassler throat). Recently
developed methods in 5d off-shell supergravity~\cite{Kugo:2000hn} might
prove useful for this analysis. The appropriate framework should be a 5d 
gauged supergravity, obtained as some flux-induced deformation of the 
AdS$_5\times T^{1,1}$ theory anticipated in~\cite{Ceresole:2000jd}.

Once the 5d theory has been understood, the next challenge will be to find
descriptions for the coupling to the SUSY breaking sector and the visible
sector. We anticipate that both the UV manifold and the $\ol{\rm D3}$ brane
will lead to a nonlinear realization of the SO(4). How precisely a D-brane 
system (containing the visible sector fields) and an anti-brane in the 
throat (comprising the hidden sector) can couple to such a nonlinearly 
realized symmetry and what the 5d description of these coupling would look
like appears to be highly non-trivial at this stage. It thus seems to be a 
long way to finally calculating soft terms.

However, some interesting questions may perhaps be raised and answered 
without knowing the details of the mechanism underlying a specific model. 
It would be interesting to see at a more quantitative level what vector 
mediation has to say about the tachyonic slepton problem of anomaly 
mediation. An important generalization of our analysis, so far conducted 
at the tree-level, would be the incorporation of loop 
effects~\cite{Gregoire:2004nn}.

\vspace{0.5cm}
\noindent
{\bf Acknowledgements}:\hspace*{.5cm} We would like to thank Carlo
Angelantonj, Emilian Dudas, Michael Ratz, Claudio Scrucca, 
Enrico Trincherini, Alberto Zaffaroni and Fabio Zwirner
for helpful comments and discussions.
M.T. thanks the Galileo Galilei Institute for Theoretical Physics
for hospitality and the INFN for partial support during the
completion of this work.

\appendix
\section*{Appendix A: A simple model for a sequestered hidden sector in the 
throat}
We take
\be
\Omega=-(T+\ol{T})+\omega^2\,(X\ol{X}-(X\ol{X})^2),\qquad 
W=W_0+e^{-T}+\omega^3 X.
\ee
in the $D=4,\;{\cal N}=1$ Lagrangian Eq.~\eqref{sugralagrangian}, so that 
its potential part becomes
\be
\begin{split}
{\cal L}\supset&\;-F_{\varphi}F_{\ol{\varphi}}\,(T+\ol{T})
-(F_T F_{\ol{\varphi}}\hc)+\omega^2\,F_\varphi F_{\ol{\varphi}}\,(X\ol{X}
-(X\ol X)^2)+\omega^2\,F_X F_{\ol{X}}\\
&+\omega^2\left(F_{\ol{\varphi}}F_X\ol{X}\,(1-2 X\ol{X})\hc\right)
-4 \omega^2\,F_X F_{\ol{X}}\,X\ol{X}\\
&+\Bigl[\bigl(3F_{\varphi}(W_0+e^{-T}+\omega^3 X)-F_T\,e^{-T}
+\omega^3\,F_X\bigr)\hc \Bigr].
\end{split}
\ee
The equations of motion read (note that those for $F_{\ol{T}}$ and 
$\ol{T}$ are unchanged from Eqs.~\eqref{fteom} and \eqref{teom})
\begin{align}
\nonumber F_{\ol{\varphi}}:\quad &F_\varphi\,(T+\ol{T})+F_T
-\omega^2\,F_\varphi\left(X\ol{X}-(X\ol{X})^2\right)
-\omega^2 F_X\ol{X}(1-2 X\ol{X})\\
&\quad-3W_0-3e^{-\ol{T}}-3\omega^3\ol{X}=0,\label{fphieom2}\\
F_{\ol{T}}:\quad&F_\varphi+e^{-\ol{T}}=0,\label{fteom2}\\
F_{\ol{X}}:\quad&\omega^2\,F_X+\omega^2\,F_\varphi X(1-2 X\ol{X})
-4\omega^2\,F_XX\ol{X}+\omega^3=0,\label{fxeom}\\
T:\quad&F_\varphi F_{\ol{\varphi}}+3F_\varphi\,e^{-T}-F_T\,e^{-T}=0,
\label{teom2}\\
\nonumber X:\quad&\omega^2\,F_\varphi F_{\ol{\varphi}}\ol{X}(1-2X\ol{X})
+\omega^2\,F_\varphi F_{\ol{X}}(1-4X\ol{X})
-2\omega^2\,F_{\ol{\varphi}} F_X\ol{X}^2\\
&\quad-4 \omega^2\,F_X F_{\ol{X}}\ol{X}+3\omega^3\,F_\varphi=0.\label{xeom}
\end{align}
As before, from Eq.~\eqref{fteom2} and the condition that the pre-uplift
superpotential in the vacuum should be $\sim\omega^2$, one can immediately 
see that $F_\varphi\sim\omega^2$. From Eq.~\eqref{teom2} it follows that
$F_T \sim\omega^2$, and from Eq.~\eqref{fxeom} we can deduce that 
$F_X\sim\omega$.

\section*{Appendix B: Alternative estimation of chiral superfield contribution}
First let us collect some supergravity relations for reference: The
kinetic function $\Omega$ and the K\"ahler potential $K$ are related by
\be
K=-3\log(-\Omega/3)\,,
\ee
where we have reinstated a factor $1/3$ suppressed in the main text. 
Given the Lagrangian of Eq.~(\ref{sugralagrangian}), we are able to deduce
the $F$ term for any chiral superfield from its equation of 
motion,
\be\label{appBf}
F^I=\frac{3}{\Omega}K^{I\bar{J}}\, (\ol{D_J W})\,.
\ee
Notice our non-standard normalization motivated by the chiral
compensator formalism.
The above calculation requires the chiral compensator $F$ term 
\be\label{appBfphi}
F_\varphi=-\frac{1}{\Omega}\left(\Omega_I F^I+3\ol{W}\right)\,.
\ee
Here $W_I\equiv\partial_I W$, $\Omega_I\equiv\partial_I\Omega$,
$K^{I\bar{J}}$ is the inverse of the K\"ahler metric 
$K_{I\bar{J}}\equiv\partial_I\partial_{\bar{J}}K$, and the K\"ahler covariant 
derivative is $D_I\equiv\partial_I+K_I$. As opposed to the main text, we 
now distinguish between upper and lower chiral multiplet indices. 

As in Sect.~4, consider a model with the K\"ahler modulus $T$, a chiral 
superfield $X$ localized at the bottom of the throat (the SUSY breaking 
sector), and a heavy chiral superfield $Y$ which couples to both $X$ and 
the visible sector. We take
\be\label{appBomega}
\Omega=\Omega_0(T,Y)+\omega^2\,\Delta\Omega(X,Y)
\ee
and
\be\label{appBw}
W=W_0+W_{\rm np}(T,Y)+\tilde W(Y)+\omega^3\,\Delta W(X,Y).
\ee
We have absorbed the superpotential contributions from all fields which have 
been integrated out in $W_0$. The non-perturbative part $W_{\rm np}$ may in 
principle involve $Y$ as well as $T$. 
In order to have a vanishing or very small 
cosmological constant we should have $W_0\sim W_{\rm np}\sim\omega^2$ including
partial derivatives, while $\Omega$ is generically of 
order one. As we will see later, the mass term $\tilde W(Y)\sim Y^2$ is 
$O(\omega^4)$ in the vacuum.

We also have $\Omega_T\sim 1$, $\Omega_X\sim\omega^2$, $W_X\sim\omega^3$, and
$W_T\sim\omega^2$. Furthermore, $K_{T\ol{T}}\sim 1$, $K_{Y\ol{Y}}\sim 1$,
and $K_{X\ol{X}}\sim\omega^2$. Hence the only term in the inverse K\"ahler 
metric that is actually warp-enhanced (rather than suppressed or 
${\cal O}(1)$) is $K^{X\ol{X}}\sim\omega^{-2}$. This is true even if there 
is mixing between $X$ and the other fields in the K\"ahler potential, as can 
easily be seen by explicitly inverting $K_{I\bar J}$.

From Eq.~\eqref{appBf} it follows that
\be
\label{appBf2}
F^{\ol{X}}\sim\omega+\tilde W_Y K^{Y\ol{X}},
\quad F^{\ol{T}}\sim\omega^2+\tilde W_Y K^{Y\ol{T}},
\quad F^{\ol{Y}}\sim \omega^2+ \tilde W_Y K^{Y\ol{Y}}.
\ee
Here we kept the leading terms in the $\omega$ as well as the yet 
unspecified $\tilde W_Y$ contributions.

We now consider the equation of motion for $Y$, replacing $F$ terms by the 
expressions given in Eqs.~(\ref{appBfphi}) and~(\ref{appBf2}). The resulting 
equation for $\tilde W_Y$ has the structure
\be
\omega^2+\tilde{W}_Y+\tilde{W}_Y^2=0\,.\label{pol}
\ee
Recall that, by assumption, the perturbative superpotential $\tilde{W}(Y)$ 
contains no linear term in $Y$ and stabilizes $Y$ at zero as long as 
non-perturbative effects and SUSY breaking are ignored. This implies that, 
in the limit $\omega\to 0$ (where these effects are switched off), 
the vacuum value of $\tilde W_Y$ vanishes. From Eq.~(\ref{pol}) it now
follows that, in this limit,
\be
\tilde{W}_Y\sim Y \sim\omega^2\,.\label{wy}
\ee
Then, we conclude from Eqs.~(\ref{appBfphi}) and~(\ref{appBf2}) that
\be\label{appBf3}
F_\varphi\sim\omega^2\,,\quad
F^{\ol{X}}\sim\omega\,,\quad F^{\ol{T}}\sim\omega^2.
\ee
At first glance one might also expect $F^Y\sim\omega^2$, but 
the leading terms in fact cancel. Indeed, returning to the equation of
motion for $Y$ (this time using the above estimates of Eqs.~(\ref{wy})
and (\ref{appBf3}) but leaving $F_Y$ unspecified), one finds
\be
F^Y\sim\omega^4\,.
\ee
Thus, the final result for the $F$ terms agrees with Eq.~\eqref{fy}.

\end{document}